\documentstyle[12pt]{article}

\newcommand{\be}{\begin{equation}}
\newcommand{\ee}{\end{equation}}
\newcommand{\bea}{\begin{eqnarray}}
\newcommand{\eea}{\end{eqnarray}}

\newcommand{\V} {{\cal V}_{n}}

\begin{document}

\begin{center}
\begin{large}
{\bf   Asymptotically Flat Holography     \\}
{\bf   and  \\}
{\bf   Strings on the Horizon  \\}
\end{large}  
\end{center}
\vspace*{0.50cm}
\begin{center}
{\sl by\\}
\vspace*{1.00cm}
{\bf A.J.M. Medved\\}
\vspace*{1.00cm}
{\sl
Department of Physics and Theoretical Physics Institute\\
University of Alberta\\
Edmonton, Canada T6G-2J1\\
{[e-mail: amedved@phys.ualberta.ca]}}\\
\end{center}
\bigskip\noindent
\begin{center}
\begin{large}
{\bf
ABSTRACT
}
\end{large}
\end{center}
\vspace*{0.50cm}
\par
\noindent
Recently, Klemm and others [hep-th/0104141]
have successfully  generalized  the Cardy-Verlinde
formula  for an asymptotically flat  spacetime  of
arbitrary dimensionality.
 And yet, from a holographic perspective, the  interpretation of this formula
remains somewhat unclear. Nevertheless,  in this  paper, we  
incorporate the implied  flat space/CFT duality  into
 a  study on    boundary descriptions
of a $d$-dimensional Schwarzschild-black hole  spacetime.    
In particular, we  demonstrate
that the (presumably) dual CFT  adopts a string-like description
and,  moreover, is thermodynamically equivalent to a string
that lives on the stretched horizon of the bulk
 black hole. Significantly, a similar equivalence 
has recently  been established  in both an asymptotically dS and
 AdS 
 context. On this basis, we argue that the asymptotically flat
Cardy-Verlinde formula does, indeed, have a holographic 
pedigree.

\newpage

\section{Introduction}

In spite of much attention, a consistent theory of quantum gravity
remains one of the most   challenging problems in theoretical physics. 
Nonetheless, one can still
hope to partially  comprehend this elusive, fundamental theory   
by looking for  imprints that it may leave  on  a semi-classical framework.  
One likely imprint is  the ``holographic principle'' \cite{tho,sus},
which indicates that a physical system of dimensionality $d$ can
be effectively described by a system that lives in one dimension
fewer.
\par
It is interesting to note that
 the holographic principle started out as
an intuitive leap of faith  from the  Bekenstein
entropy bound \cite{bb}.  Since its literary inception, however,  this
principle has attained
 considerable support; in particular, 
by way of the anti-de Sitter (AdS)/conformal field theory (CFT)
correspondence \cite{mal,gub,wit}. The essence of this  correspondence
is that a $d$+1-dimensional theory of AdS gravity (i.e., Einstein
gravity with a negative cosmological constant)  is dually
related to a $d$-dimensional CFT that lives on a timelike
boundary of the  AdS bulk. For instance,  boundary values
of bulk fields serve as sources for the corresponding CFT. This realization
has lead to explicit calculations of  boundary correlation
functions in terms of strictly bulk formalism \cite{agmoo}.  
\par
On the basis of the holographic principle, it
has been argued  that  some sort of analogue to the AdS/CFT duality
 should exist for
any  bulk theory that includes gravity \cite{sw}. However, this is 
anything but a trivial extension, considering 
that gravitational theories with
a non-negative cosmological constant have no accessible boundary at spatial
 infinity.  At least naively, the 
existence of such a boundary is  critically
important to many relevant calculations  
on account of the infrared-ultraviolet
connection \cite{sw}.  That is to say, the  high-energy regime
of  a boundary-induced  CFT can only be accessed in the large
radial limit of the dual bulk theory.  Such 
accessibility is important because as
one moves deeper and deeper into the bulk (i.e., towards smaller radii),
the CFT probes  lower and lower energies, and  the fine-grained
details of the holographic picture are lost.  
\par
In spite of the above discussion, an analogous de Sitter (i.e., positive
cosmological constant)  or dS/CFT
correspondence  
has indeed been established \cite{str2}.\footnote{For further
citations in support of the  dS/CFT duality, see Ref.\cite{med}.
Also see Refs.\cite{zmed,zmyung,zdas}.}
The problem of a missing boundary at spatial infinity
has nicely been  circumvented  with the assumption that the
dual CFT  is now  a Euclidean one that lives on a spacelike
boundary of the dS bulk. It should be noted, however, that
the dS/CFT duality remains on weaker ground than its AdS
analogue. This can primarily be attributed to dS spacetimes  lacking 
a string-theoretical description and an objective observer \cite{deh}.
\par
Also of interest is the flat-space limit of the AdS/CFT correspondence,
which   translates into a bulk gravity theory with a vanishing cosmological
constant. In principle,  this  should be a perfectly valid
limiting procedure.  In practice, however, rigorous calculations
are inhibited by  a non-localization of the holographic
mapping as the flat-space  limit is approached \cite{sussy}.
\par
An interesting subplot in the AdS/CFT  correspondence
is the renowned Cardy-Verlinde formula \cite{ver}.
In the cited paper (also see \cite{sav}), Verlinde  
exploited the  relevant
duality to demonstrate that, for an arbitrary-dimensional bulk,
 the  thermodynamics of the dual CFT 
 could be  expressed in a form which mimics
the Cardy entropic formula \cite{car}.  Also of interest, the
sub-extensive contribution to the CFT entropy plays
the role of the Cardy ``central charge''.  
Paradoxically, the Cardy
formula  originated in  a strictly two-dimensional (CFT) framework, 
and there is
 no {\it a priori}
basis for its  generalization   to  higher-dimensional theories.
Furthermore, the Cardy central charge is
a   Virasoro algebraic parameter; with the Virasoro algebra 
 describing the symmetries of strictly two-dimensional CFTs \cite{vir}. 
\par
Let us briefly point out that the Cardy-Verlinde formula \cite{ver} 
has since been
generalized to a myriad of scenarios; including
 asymptotically AdS (for instance, \cite{cai}),
asymptotically dS (for instance, \cite{dan}) and, even, asymptotically
flat \cite{klemm,jing} spacetimes.\footnote{For  an up-to-date 
list of citations,
see Ref.\cite{youm}.} Moreover, the status of the ``Casimir entropy''
(i.e., the sub-extensive contribution to the CFT entropy)  
as a  legitimate analogue to the Cardy central charge has since  been
substantiated  \cite{klemm,zklemm,hal9000}. In particular, these studies 
have identified the Casimir entropy  with
 the ``C-function'' of  a renormalization-group flow (when
the conformal symmetry of the boundary theory is broken),
in perfect analogy to the central charge of a two-dimensional broken CFT
\cite{zamo}.
\par
Let us return our attention to the paradox of 
 CFT entropies consistently adopting  Cardy-like descriptions,
regardless of the dimensionality. 
In a sense, this is not particularly surprising in view of some prior
studies by Carlip and Solodukhin \cite{carlip,sol}. These authors
have shown that the two-dimensional Cardy formula is relevant to the
entropic calculation of virtually any spacetime (of any dimensionality) that
admits a black hole (or black hole-like) event horizon. 
In spite of this universality,  a lack of clarity at the interpretative
level remains conspicuously absent.\footnote{See Ref.\cite{crev} for relevant
 discussion.} Recently, however, Halyo  was able to shed some
light on this matter \cite{xhal}.
 In the context of a $d$+1-dimensional asymptotically
dS bulk, this author has shown that the dually related ($d$-dimensional) CFT   
behaves, thermodynamically, just like a string.
 In particular, a linear relationship between the     entropy
and  extensive energy of the boundary theory is in evidence.
In view of  this  stringy description,
 the $d$-dimensional CFT  can 
be interpreted  as having only two relevant dimensions.
\par
Halyo went on to demonstrate \cite{xhal} 
that the CFT thermodynamic properties, including
an effective string tension and energy, are directly related to those of a
string that lives on the ``stretched'' (cosmological) 
horizon\footnote{A stretched black hole (or
cosmological) horizon \cite{xsus,xstu} 
 refers to a timelike surface that lies at a distance of 
the fundamental string scale ($l_s$) above the event horizon.}
of the asymptotically dS bulk. (This latter picture follows
from the ``black hole-string correspondence'', as  originally
advocated  
by Susskind \cite{xsus}.) Furthermore, when the properties
of the stretched  horizon have  properly been renormalized (see the end
of this section), 
then these two stringy descriptions
are, in fact, thermodynamically  equivalent.  
That is, the CFT can alternatively  be viewed
 as a horizon-based string having  a renormalized tension.
Notably, this program was later generalized
for the case of an  AdS-black hole bulk \cite{paper}. 
Given that a string is fundamentally a two-dimensional object, a Cardy-like
formulation  for the entropy of any CFT or, by duality, any black
hole begins to make sense. 
\par
The focus of the current paper is a further generalization of the  Halyo
treatment \cite{xhal} (also see \cite{paper})  
to the case of an asymptotically
flat black hole spacetime.  Such a study has importance for
the following two reasons. Firstly,  considering
 the priorly discussed work of Carlip and Solodukhin \cite{carlip,sol},
one might anticipate that a stringy description 
 of  a duality-induced CFT 
 is a universal feature of any bulk theory that admits an event horizon.
Thus, for completeness sake, it is important to verify that this
description  persists in the case of an asymptotically flat bulk.
\par
Secondly, let us first point out that a  definitively 
 dual  CFT  for   an 
 asymptotically flat spacetime remains
a relatively  open question \cite{sussy}.  Nonetheless,  
Klemm {\it et al}. \cite{klemm}  have  
 managed to generalize the Cardy-Verlinde formula 
for  just such a flat-space
 scenario.\footnote{Jing \cite{jing} has also considered
such a generalization.}
   However, it remains unclear, at least
to the current author, if this generalized Cardy-Verlinde formula  should
be interpreted as an intriguing coincidence 
or as a  true manifestation of the holographic principle. (In this regard,
we note that the formula in question, although  certainly valid,
did not follow directly from  holographic  
arguments.\footnote{This is contrary
to the original Cardy-Verlinde expression \cite{ver,sav} and many
of its subsequent
generalizations.} We further note that the associated Casimir energy
lacks the monotonicity properties, 
as relevant to a renormalization-group
flow, which are evident in  AdS-inspired dualities \cite{klemm,zklemm}.)
 We believe that this issue can be 
somewhat clarified  by  way of the 
``CFT-stretched horizon  correspondence''.
That is to say, if such a duality is in evidence for an asymptotically
flat bulk, as  it is  for  
asymptotically AdS and dS scenarios \cite{xhal,paper},
it would strongly support the holographic  viewpoint.     
\par
\par
The remainder of this paper is organized as follows. In Section 2,
we begin by introducing the bulk solution of interest; 
namely, an arbitrary-dimensional
Schwarzschild (i.e., asymptotically flat) black hole.
 The relevant black hole thermodynamics are presented, and
these are used to formulate the thermodynamics of a (presumably)
dual CFT.  It is then shown that the CFT entropy  satisfies
  Klemm {\it et al.}'s   generalization \cite{klemm}
of the Cardy-Verlinde formula \cite{ver}.
We  also  demonstrate the anticipated string-like behavior
of the CFT \cite{xhal,paper} and accordingly identify an 
effective string tension
and energy.
\par
In Section 3, we consider a much different effective theory; with
this one living  on a boundary close to the horizon of the
Schwarzschild-black hole.
Following  Susskind's proposal of a black hole-string 
correspondence \cite{xsus}, we demonstrate that the near-horizon 
geometry can  be described as a  single fundamental
string that lives on this so-called  stretched horizon.  
Furthermore, the associated thermodynamic properties of the string
are appropriately identified.
\par
In Section 4, we go on to compare  these two  stringy
descriptions  (of the Schwarzschild bulk)  and ultimately
confirm their  thermodynamic equivalence.
Essential to this identification is a coordinate rescaling
of the properties of the string at the stretched horizon.
This process  can be viewed as a renormalization that accounts
for the gravitational red shift occurring between the horizon
and a  cosmological observer (see below). 
Finally, Section 5 contains
a summary and an overview.  
\par
Before proceeding to the main body of the paper, we present a brief
discussion on the black hole-string correspondence,
as this duality plays a prominent role in the analysis to follow.
\par
Roughly a decade ago, Susskind  proposed \cite{xsus} a  one-to-one
correspondence between sufficiently massive
 black holes and highly excited fundamental strings.
Let us now review the logistics of this proposal.   
As string coupling increases, the size of a string state
must ultimately fall below its Schwarzschild radius and,
hence, the string state should naturally  evolve into a black hole
\cite{xtho}.
 Conversely,
as string coupling decreases, the size of a black hole  must eventually
fall below the string scale and, hence, the near-horizon geometry
should no longer be interpreted as a black hole {\it per se}.  Thus, 
a black hole and some sort of string configuration 
can be regarded as strongly
and weakly coupled manifestations of the same entity.  Now consider
that, at large
enough mass, a typical string state consists of a small
number of highly excited strings. However, as such a  configuration
approaches the high-temperature conditions of the  horizon (i.e., the Hagedorn
temperature),
a single string  will become the entropically preferred state \cite{aaw}.
Alternatively,  an external observer  would view the various strings 
as having  melted
together into a single string that  fills up the
stretched horizon \cite{sss}.  On the basis of these arguments, 
 a one-to-one 
correspondence between black holes and strings does indeed follow.
\par
Contrary to the above perspective, there is, of course, an observed 
discrepancy
between black hole and string thermodynamics. 
In particular,  the entropy of a string  has a  linear energy dependence,
whereas  the entropy of (for instance) a  four-dimensional 
Schwarzschild black hole depends quadratically on its mass.
 However, Susskind  has conjectured \cite{xsus}  
 that this discrepancy
can be accounted for with an appropriate renormalization of the string energy. 
  Such a renormalization is, indeed, necessary  by virtue of
the large gravitational red shift  occurring between the stretched horizon
and an external observer. Exploiting the Rindler-like 
 near-horizon geometry of a Schwarzschild black hole,
Susskind was able to  demonstrate the feasibility of this conjecture.
\par
The validity of the  black hole-string  correspondence  has 
since been rigorously substantiated  for 
various  black hole (as well as  black hole-like) scenarios 
\cite{xhrs,xhkrs,xhp,xhal2,xdv,xdam,xugk}.  Generally speaking,
 the counting of string states yields an entropy that coincides 
with  the anticipated Bekenstein-Hawking  form up to some
ambiguity in the numerical coefficient.   Let us further note 
that the same  philosophy has also been applied to special
classes of supersymmetric extremal and near-extremal black holes.
 (For instance, Ref.\cite{vaf} and, for a review,  Ref.\cite{xpeet}.)
Remarkably, in these cases, the Bekenstein-Hawking form
has been reproduced with no ambiguity.  Significant to
this precision is a special property of such models;
namely, supersymmetry protects the mass
against renormalization \cite{xsen}.

\section{A CFT Description of Schwarzschild Black Holes}

In this section, we will review a proposal by Klemm {\it et al}.
 \cite{klemm} of a generalized Cardy-Verlinde
formula \cite{car,ver} for an asymptotically flat black hole.
We will also re-interpret the dual CFT (as implied
by the generalized formula)  as an
effective stringy theory in analogy to Refs.\cite{xhal,paper}. 
\par
Let us begin here by considering the bulk model of interest;
 namely, a spherically symmetric 
 black hole  in an
 $n$+2-dimensional spacetime with    vanishing
cosmological constant.\footnote{In this study, we will limit 
considerations to $n\geq 2$, as it is well established that 
an  asymptotically flat, three-dimensional spacetime does 
not admit black hole solutions. We are also assuming no conserved
charges other than the mass.}  
In a  static gauge,   such a black hole
can be uniquely  described by the well-known Schwarzschild solution
(generalized for arbitrary dimensionality). That is: 
\be
ds^2_{n+2}=-h(r)dt^2+{1\over h(r)}dr^2+r^2d\Omega^2_{n},
\label{1}
\ee
where:
\be
h(r)=1-{\omega_{n}M\over r^{n-1}},
\label{2}
\ee
\be
\omega_{n}={16\pi G \over n \V}.
\label{3}
\ee
Here, 
$d\Omega_n^2$
denotes the line element of a
 unit $n$-sphere with volume
 $\V$, $G$ is
the $n$+2-dimensional Newton constant, and $M$  
 (a constant of integration)  is the
 conserved mass of the associated black hole. 
\par
For a non-vanishing (positive) $M$, there will be a single black hole
horizon,  which corresponds to the  positive root of $h(r)$.
Denoting this horizon by $r=R$, we thus obtain:
\be
 R^{n-1}= \omega_{n} M.
\label{4}
\ee
\par
The associated black hole thermodynamics (energy, temperature, entropy) can  
readily be  identified via standard techniques \cite{gh2}:
\be
E_{Sch}=M,
\label{5}
\ee
\be
T_{Sch}={1\over 4\pi} \left.{dh\over dr}\right|_{r=R}= {n-1\over 4\pi R}, 
\label{6}
\ee
\be
S_{Sch}= {\V R^n\over 4 G}={4\pi\over n\omega_n}R^n.
\label{7}
\ee
\par
Following Klemm {\it et al}. \cite{klemm}, let us assume that
Schwarzschild black holes, like their  AdS-Schwarzschild  counterparts,
have the following dual description:  
an $n$+1-dimensional CFT that lives  on a timelike boundary of
the bulk spacetime. 
The CFT thermodynamics should then be related to
those of a bulk up to a simple red-shift factor (with this ambiguity being
a direct consequence of the conformal symmetry of the boundary theory)
\cite{mal,gub,wit}.
Because of the relative simplicity of the bulk theory, we
will assume  that, in this case, the red-shift factor is equal to
 unity. Significant
to this choice is the existence of a single length scale, $R$,  for the bulk 
geometry. (Actually, this choice is difficult to justify;  except that,
because of the conformal symmetry at the boundary, 
we are free to do so. One might say
that we started with the generalized Cardy-Verlinde formula \cite{klemm}
and then defined the CFT  thermodynamics so as to obtain a suitable match.
The significance of this distinction will be elaborated on 
in the final section.)
\par
Given the above considerations, we have:   
\be
E_{CFT}= E_{Sch}= {1\over \omega_{n}}R^{n-1},
\label{8}
\ee
\be
T_{CFT}=T_{Sch}=  {n-1\over 4\pi R}, 
\label{9}
\ee
\be
S_{CFT}=S_{Sch}= {4\pi\over n\omega_n}R^n,
\label{10}
\ee
where we have applied Eq.(\ref{4}) in obtaining the first result.
Note that the  CFT  entropy  is, in fact, uniquely defined, as entropy
 is  never 
affected by the 
choice of scale 
\cite{wit}.
\par
Now  following Verlinde \cite{ver},
we will  define the Casimir contribution ($E_C$) to the  CFT energy  
as the violation of  $E(\lambda S,\lambda V)=\lambda
E(S,V)$; i.e., the violation in the Euler identity. 
This definition  leads to the following relation \cite{ver}
(now suppressing the subscript CFT on relevant thermodynamic quantities): 
\be
E_C= n\left[E+pV -TS\right],
\label{10.66}
\ee
where $V=R^{n}\V$ is the volume enclosed by the CFT boundary and
 $p=E/nV$ is the pressure. Here, we have  
assumed  the equation of state 
for radiative matter, as is appropriate for a theory with a traceless
stress tensor. 
Straightforward evaluation of the above   yields:
\be
E_C= 2{R^{n-1}\over \omega_n}= 2E.
\label{10.6}
\ee
\par
Let us recall the Cardy-Verlinde formula for
the CFT entropy as relevant to an asymptotically AdS bulk \cite{ver}:
\be
S={2\pi R\over n}\sqrt{E_C\left[2E-E_C\right]}.
\label{cvold}
\ee
Naively applying this formula to the asymptotically flat case,
we obtain the nonsensical result of $S=0$. However, this 
unpleasant situation  can be  dealt with by first noting 
that the standard Cardy formula for a two-dimensional 
CFT \cite{car}:\footnote{Note that $c$ is the central charge
and $L_{o}$ is the eigenvalue of the zero-mode generator of
the corresponding Virasoro algebra, which can be used
to describe the symmetries of a two-dimensional CFT \cite{vir}.} 
\be
S=2\pi\sqrt{{c\over 6}\left[L_{o}-{c\over 24}\right]}
\label{cardy1}
\ee
takes on a simplified form when the conformal weight of the ground
state is zero. This simplification being:
\be
S=2\pi\sqrt{{cL_{o}\over 6}}.
\label{cardy2}
\ee
By direct analogy, this consideration
 implies that, under  ``appropriate circumstances'',
the Cardy-Verlinde should be modified as follows:
\be
S={2\pi R\over n}\sqrt{E_C(2 E)}.
\label{cv}
\ee
\par
It is not difficult to verify that  this  revised form
of the Cardy-Verlinde entropy (\ref{cv}) 
is, indeed, satisfied by the  thermodynamic properties 
(\ref{8},\ref{10},\ref{10.6}) of  our
conjectured CFT.
Furthermore,
it  is  straightforward to substantiate the following relation:
\be
S={4\pi R\over n}E.
\label{11}
\ee
\par
The significance of the above expression is the direct proportionality
that  exists between the CFT entropy and  energy.  (It
is perhaps a subtle distinction, but, to a boundary observer,
$R$ represents a fixed  ultraviolet cutoff \cite{sw} rather than
a mass-dependent horizon radius. That is to say, this string-like behavior
is not evident to a bulk observer; even though, strictly speaking,
the thermodynamic relations are equivalent.)
Notably,  such
 linearity is  a characteristic behavior  of strings \cite{xpeet}. 
 With this in mind, we can re-interpret Eq.(\ref{11})
as the equation of state for an ``effective string'' having an energy: 
\be
{\cal E}={2C\over n}E
\label{11.6}
\ee
and having a tension:
\be
{\cal T}={C^2\over 2\pi R^2}.
\label{11.3}
\ee
Here,  we have introduced  a yet-to-be-determined parameter, $C$.
On the basis of prior studies \cite{xhal,paper},
it can be anticipated  that, in general,  $C=C(R;n,k)$.
\par
The above string-like behavior is  not a trivial outcome, inasmuch as 
 the CFT
 typically has at least three spacetime dimensions, whereas
a string is  fundamentally a two-dimensional entity.
This outcome is, however, not a complete surprise,
considering that
the relevant degrees of freedom for  {\it any} black
hole can evidently be described by a two-dimensional CFT 
\cite{carlip,sol}. That the 
 Cardy-Verlinde description itself \cite{ver}  has 
 its genesis in two dimensions \cite{car}  also supports this notion.
It would appear that some universal but as-of-yet-unknown principle underlies
these various approaches.

\section{A Stringy Description of Schwarzschild Black Holes}

In the preceding  section, we  have shown that the
 CFT of interest (i.e.,  a CFT that is possibly dual to an asymptotically flat
spacetime)  exhibits  some  characteristics of a string-like object.
With this mind, we will next consider a 
 stringy description of a  Schwarzschild black hole   
as based on a one-to-one
 correspondence
that appears to exist between black holes and strings \cite{xsus}. 
More specifically,
Susskind has proposed that a massive  black hole  can   be
identified with  a highly excited string which is located
 at  the so-called 
stretched horizon \cite{xstu,sss}.  Furthermore, Susskind
has argued that the apparent thermodynamic differences (between 
these two pictures) 
can be attributed to the long-range effects of
the  gravitational field; that is, the immense red shift  occurring
  between the stretched horizon and a distant
observer. (See Section 1 for supplementary discussion.) 
\par
To exploit the proposed correspondence, it is first necessary
(or, at least, convenient)
to transform the near-horizon  black hole geometry
into   a Rindler-like form. We can accomplish this task by
introducing  a new radial coordinate, $y$, in accordance with
$r=R+y$ (with $y<< R$  being assumed). Up to first order in $y/R$,
the metric function $h(r)$ (\ref{2}) can now be written as: 
\be
h(y)\approx {n-1\over R} y. 
\label{12}
\ee
\par
Incorporating  the above into Eq.(\ref{1}), we find the 
following near-horizon form
for the  line element:
\be
ds^2_{NH}=-y{n-1\over R}dt^2+
{1\over y}{R\over n-1}dy^2+R^2d\Omega^2_{n}.
\label{13}
\ee
\par
It is useful if $y$ is replaced with a coordinate that
directly measures the proper radial distance  from 
the horizon.
Denoting the proper distance as $\rho$, we  obtain
 (up to the first perturbative
 order):
\be
\rho\approx \sqrt{{R\over n-1}}\int^{y}{dy\over\sqrt{y}}
=2\sqrt{{R y\over n-1 }}.
\label{14}
\ee
The near-horizon line element (\ref{13})
can now be expressed in the following Rindler-like form:
\be
ds^2_{NH}=-{(n-1)^2 \over 4 R^2}\rho^2 dt^2+
d\rho^2+R^2d\Omega^2_{n}.
\label{15}
\ee
\par
The  near-horizon  geometry can  be identically formulated as a 
  Rindler
spacetime:
\be
ds^2_{NH}=-\rho^2 d\tau^2+
d\rho^2+R^2d\Omega^2_{n},
\label{17}
\ee
provided that the associated temporal coordinate is defined  as follows:
\be
\tau\equiv {n-1\over 2R} t.
\label{16}
\ee
Note that the Rindler time,  $\tau$, is a  dimensionless quantity. 
\par
Next, let us  consider the thermodynamics as measured by a hypothetical
 Rindler observer at the stretched horizon. The dimensionless horizon
 temperature
is  known to be \cite{big}:
\be
T_R={1\over 2\pi}.
\label{18}
\ee
Meanwhile, the  Rindler entropy should still be given by the 
Bekenstein-Hawking area law \cite{bek,haw} (or its
$n$+2-dimensional analogue), 
 and so:
\be
S_R=S_{Sch}.
\label{19}
\ee
\par
To determine the  dimensionless Rindler energy, we simply apply the first
law of thermodynamics; that is, $dE_{R}=T_R dS_R$. 
This process yields:
\be
E_R={1\over 2\pi}S_R={2R^{n}\over n \omega_n},
\label{20}
\ee
with the right-most relation following from Eqs.(\ref{7},\ref{19}).
\par
Significantly,
the linear relation between Rindler entropy and energy  is (once again) 
indicative
 of  string-like behavior. By way of Susskind's conjecture 
\cite{xsus},
we can attribute this  behavior to the state of a   single string 
 that lives on  the stretched
horizon. Note that  the associated  string tension, in
dimensionless Rindler
coordinates, can readily be identified as:
\be
{\cal T}_R={1\over 2\pi}.
\label{ten}
\ee
Meanwhile, the dimensionless string energy is trivially ${\cal E}_R=E_R$.
\par
For a hypothetical observer located near the  horizon,
the
 fundamental length scale will be the string length, $l_s$,
as it is this quantity that, by definition, determines the radial extent of
the stretched horizon \cite{xstu,sss}. 
 Hence, we can obtain the
dimensional thermodynamics of the stretched horizon by 
 rescaling  the
Rindler relations  in terms of  this length.  In particular, quantities
with units of energy (or inverse length) should be divided by $l_s$.
That is: 
\be
T_{SH}={T_R\over l_s}={1\over 2\pi l_s},
\label{22}
\ee
\be
S_{SH}=S_R= S_{Sch},
\label{huh}
\ee
\be
{\cal E}_{SH}= {{\cal E}_R\over l_s}= {2R^{n}\over n \omega_n l_s},
\label{23}
\ee
\be
{\cal T}_{SH}={{\cal T}_{R}\over l_s^2}={1\over 2\pi l_s^2}.
\label{21}
\ee
It is interesting to note that $T_{SH}$  agrees with  
the Hagedorn temperature
of a string \cite{xpeet}.
\par
Let us re-emphasize that, in accordance with the black hole-string 
correspondence \cite{xsus},
the apparent discrepancy in thermodynamic properties  (between
those of  the stretched horizon and those of the  black hole)
  should
be attributable to the long-range effects of the  gravitational field.
Notably, this conjectured duality has, indeed, been verified for
a number of models (for instance, \cite{xhkrs}) up
to some ambiguity in the  numerical coefficients.
Furthermore,    it has been demonstrated that the same arguments lead
to a thermodynamic identification between the
string at the stretched horizon and the (effective) string on the CFT boundary
\cite{xhal,paper}. This last point is of particular
relevance to the analysis of the following section.

\section{A CFT/String Correspondence?}

In the preceding analysis,  we have observed two different
string-like descriptions of an asymptotically flat (black hole) spacetime;
with the strings
in question being located on   the ``CFT  boundary''   and the stretched black
hole horizon. The purpose of the current section is to ascertain
if these stringy theories are thermodynamically equivalent.
Given that  such an equivalence  has been found for  both
 asymptotically AdS \cite{paper} and dS \cite{xhal} bulk
 geometries, we will view a similar outcome here as strong support
for the proposed duality between asymptotically flat spacetimes and
 CFTs \cite{sussy}.
\par
The stringy theories of interest can only be judiciously compared
 if  the pertinent measurements are being carried out by a common
observer.  Hence,  
 it is  necessary that the thermodynamics 
of the stretched horizon be suitably  rescaled for a 
cosmological
observer (i.e., one  who is  far away  from the horizon
in terms of all relevant length scales).  We  can accomplish this
task  by 
 rescaling  the  (dimensionless) Rindler
thermodynamic quantities (\ref{18}-\ref{ten})
 in terms of the ``cosmological'' time coordinate, $t$.  Utilizing
the precise relation between $t$ and the  
   Rindler
time coordinate  ({\it viz} Eq.(\ref{16})),  we thus  obtain
the following rescaled quantities:
 \be
T_{SH}^{\prime}={d\tau\over dt}T_{R}={n-1\over 4\pi R},
\label{24}
\ee
\be
S_{SH}^{\prime}=S_{R}=S_{Sch},
\label{26}
\ee
\be 
{\cal E}_{SH}^{\prime}={d\tau\over dt}{\cal E}_{R}= {n-1\over n \omega_{n}}
R^{n-1},
\label{25}
\ee
\be
{\cal T}_{SH}^{\prime}=
\left({d\tau\over dt}\right)^2 {\cal T}_{R}={(n-1)^2
\over 8\pi R^2}.
\label{27}
\ee 
Note that  a prime indicates that a rescaling has taken place. 
\par
As an aside, let  us  compare the above outcomes
with those of Eqs.(\ref{22}-\ref{21}); that is, the
thermodynamic properties of the string at the stretched horizon as
measured by a local observer.
In going from a horizon to a cosmological perspective,
we observe  an  effective renormalization of 
roughly $\l_s^{-1}\rightarrow R^{-1}$.
Since the black hole-string correspondence \cite{xsus} presumes  that 
$R>>l_{s}$, this renormalization implies
 a significant reduction in the energy, string tension and
temperature as measured by an external observer.  Such
a renormalization is, however, expected
by virtue of the large gravitational red shift
occurring between the horizon and a cosmological 
vantage point \cite{xsus}. 
\par
In the prior related works \cite{xhal,paper},
a further rescaling was necessary  to 
account for  an observer who is specifically  located at the CFT boundary.
We happily note that this is not the case here, since the 
relevant red-shift factor (between the bulk and the CFT boundary)
was originally taken  to be unity.
Thus, we are now in a position to directly compare
the  thermodynamic properties of the 
stringy CFT (\ref{9},\ref{10},\ref{11.6},\ref{11.3})
with the renormalized properties of the stretched horizon (\ref{24}-\ref{27}).
This comparison goes as follows.
\par
We begin here by noting that the entropies $S$ and $S^{\prime}_{SH}$
trivially agree, as
  this dimensionless quantity
is unaffected by any coordinate rescalings.  On a less trivial note,
we observe the  following thermal coincidence (cf. Eqs.(\ref{9},\ref{24})):
\be
T= {n-1\over 4\pi R}=
T_{SH}^{\prime}.
\label{28}
\ee
\par
Next, let us  compare the pair of expressions for the string tension.  
Recalling Eq.(\ref{11.3}), we see that the CFT-inspired tension
depends on an arbitrary parameter, $C$. Our approach 
will be to identify ${\cal T}$ with ${\cal T}_{SH}^{\prime}$ (\ref{27})
and use this relation to define $C$.  (This  {\it ad hoc}
identification will be suitably be tested  via the string-energy comparison
to follow.) On this basis, we find: 
\be
C= {n-1\over 2}.
\label{31}
\ee
It is interesting that $C$ is strictly a (dimensional-dependent) constant.
This is in contrast to studies in an AdS and a dS bulk context,
where the analogous parameter was found to depend directly on  the
horizon radius \cite{xhal,paper}.
\par
Finally, let us consider a comparison of the string energies.
Substituting the above identity for $C$ into Eq.(\ref{11.6})
for the CFT-inspired string energy, we obtain:
\be
{\cal E} = {n-1\over n \omega_n} R^{n-1},
\label{what}
\ee
where we have also incorporated Eq.(\ref{8}).  By way of
the above and Eq.(\ref{25}), it is now clear to see that:
\be
{\cal E}={\cal E}^{\prime}_{SH},
\label{who}
\ee
thus  establishing the thermodynamic equivalence of
the two stringy theories.
\par
To briefly summarize, we have demonstrated a precise correspondence between the
renormalized  thermodynamics of  the stretched horizon and the
thermodynamics of  the (conjectured)  conformal boundary theory. 
This  duality suggests
that the string which appears to live on the CFT boundary 
(cf. Eqs.(\ref{11.6},\ref{11.3})) is
really just the string at the stretched horizon as viewed
by a cosmological  observer.  Moreover, the observed correspondence
adds support to the notion  of a  conformal boundary
description of an asymptotically flat spacetime (given that
analogous outcomes were obtained \cite{xhal,paper} with  bulk spacetimes
for which a CFT duality is well-established). 
  The relative simplicity of 
the prior calculations
may  obscure the non-triviality of this  outcome.  Hence, let us emphasize
that this pair of boundary theories  has  no
 {\it a priori} relationship. 
\par
As a final note in this section, we point out  an  agreement (up to
an insignificant  constant factor)
between the ``stringy energy'' (${\cal E}$) and the
total energy ($E$) of the CFT. This may seem a trivial outcome;
however, no such agreement was found  in the prior, related studies
in an AdS and dS context \cite{xhal,paper}.   For instance, given an
 AdS-Schwarzschild bulk spacetime, these two energies
were found to differ by $\approx E_E-2E_C$ (where $E_E$ is
the purely extensive contribution to the energy of
the CFT).
Considering that this energy discrepancy
was attributed to an ambiguity in defining conserved charges
in an asymptotically AdS or dS  background \cite{bal,newbal},
it is quite encouraging that no such discrepancy arises
in asymptotically flat spacetime (where conserved
charges are unambiguously defined).

\section{Conclusion}

In summary, we have been studying some  holographic aspects of 
 an  asymptotically 
flat spacetime of arbitrary dimensionality. In particular, we have  
focused
on static, spherically symmetric  bulk solutions admitting a single
black hole horizon;  that is,
$n$+2-dimensional Schwarzschild black holes.  
 The analysis formally began with  a review of
the pertinent bulk formalism, including  the black hole
thermodynamic expressions.  
After these preliminaries,   we considered   a pair of effective
boundary theories; both of which have  a conjectured   duality 
with  the   black hole bulk  spacetime.
 (We  again point out that our  treatment follows
from a related paper by
  Halyo \cite{xhal}. Also see Ref.\cite{paper}.) 
\par
The first boundary theory of interest was   a conformal
field
theory that lives on a timelike  hypersurface  of the  asymptotically
flat spacetime.
Such a  CFT is expected to be dually related to the bulk theory,
presuming a well-defined flat-space limit 
of the AdS/CFT correspondence \cite{mal,gub,wit,sussy}.
On the basis of this  proposed duality and a relevant study by
Klemm {\it et al}. \cite{klemm}, we
were able to identify  various   thermodynamic properties of the CFT and
show that they satisfy a 
Cardy-Verlinde-like form \cite{car,ver}.
Moreover, we were able to  demonstrate a linear 
relationship   between the entropy and
 energy of the CFT. 
Significantly, this outcome  
 implies that the CFT can also be  interpreted
as a  string that lives on the associated cosmological boundary. 
On the basis of  this stringy description, we identified
  an effective string tension and energy for the CFT.
\par
The second  boundary  theory  under consideration was based on
 one-to-one  correspondence that  has been conjectured 
 between  massive black 
holes    and  highly excited strings  \cite{xsus}.
 This  duality is most  evident when one
transforms the near-horizon geometry of a black hole into a Rindler-like
form.  Following this program, we are able to deduce
the thermodynamic properties of a  string that lives on
the (so-called) stretched horizon of a Schwarzschild black hole.
 Although
the stringy thermodynamic relations  appear to differ 
from those of the associated   black hole, it has  convincingly been
argued (for instance, \cite{xhkrs})
that  such  discrepancies 
 can be  attributed to the long-range effects of the gravitational
field.    
\par 
In the final portion of this analysis,
we  compared the thermodynamic relations for this
pair of effective  boundary theories. For such  a comparison,
it was first necessary  to rescale  the thermodynamic properties of
the stretched horizon.  In particular,
we invoked a  coordinate transformation from Rindler time to 
``cosmological'' time. In this way, both boundary theories could be viewed
from the perspective of a common cosmological observer. It
is interesting to note that the net effect
of this rescaling is  a significant renormalization
of the tension, temperature and energy of the string
at the stretched horizon.
 Such a  renormalization
can be attributed to the substantial  gravitational red shift 
that arises between the horizon and a cosmological 
observation point \cite{xsus}.  
\par
Ultimately, we were able to demonstrate a precise equivalence between
 the renormalized thermodynamics of
the stretched horizon and the thermodynamics of the ``stringy CFT'' 
(provided   that an arbitrary parameter had been 
suitably fixed). These  identifications included the temperature,
string energy, string tension and, trivially, the entropy.
This remarkable equivalence implies that  the
CFT thermodynamics can   be re-interpreted as 
the renormalized thermodynamics of a string
that lives nears the black hole horizon. 
Given that the two boundary theories are  {\it a priori} unrelated,
we find this identification to be   considerably non-trivial. 
\par
 In conclusion,  the  results of this paper appear to  be
particularly significant on two distinct levels; let us  highlight these 
   in turn.  
Firstly, we again point out  the success of  this general
treatment   for  asymptotically dS \cite{xhal},
asymptotically AdS \cite{paper} and, now, asymptotically  flat bulk scenarios.
Hence, we are 
 suitably positioned  to  extrapolate  a universal link 
between  CFTs on  asymptotic boundaries  and  the thermodynamics
of  horizon-based strings. Ultimately, this link may help us understand
why any relevant  CFT entropy consistently  conforms to a Cardy-like
description \cite{car,ver}.  
In this regard,  it is interesting to take note of
some  prior  studies by Carlip and Solodukhin \cite{carlip,sol}. 
These authors demonstrated
that a  Cardy-like description of  entropy 
is likely a universal characteristic of black hole (and black hole-like)  
geometries.  Since all paths lead to the Cardy formula,
one would expect that these various approaches are probing
the same fundamental principle. Unfortunately, this principle
may  remain  obscured (at least in the short run) 
by our   semi-classical limitations 
in understanding gravity. 
\par
Secondly, let us consider  our  results in the context of
the proposed flat space/CFT correspondence. Here,
we again point out  that our CFT thermodynamic framework was based,
in large part,  on a generalization of
the Cardy-Verlinde formula for an asymptotically flat spacetime
({\it viz}. Klemm {\it et al}. \cite{klemm}). That is to say, we first
assumed the formula and then utilized  it in defining
 the CFT thermodynamic properties.
This is converse to the typical  practice  whereby one
exploits the relevant duality (AdS/CFT or dS/CFT) 
in deriving the corresponding CFT  thermodynamics.
Thus,  the holographic interpretation of the flat-space
Cardy-Verlinde formula is somewhat unclear. However, 
 the results of this paper would appear to  support
a holographic pedigree for this formula, given that
similar outcomes have previously been  achieved in scenarios having  
well-established dualities \cite{xhal,paper}. In spite
of our  ``success'',   it would still be preferable
to have this  holographic connection    
 established on a more rigorous level. Hopefully,
this issue will be addressed in a future investigation.

\section{Acknowledgments}
\par
The author  would like to thank  V.P.  Frolov  for helpful
conversations.


\begin{thebibliography}{99}

\bibitem{tho} G. 't Hooft, ``Dimensional Reduction in Quantum Gravity'',
gr-qc/9310026 (1993).
\bibitem{sus} L. Susskind, J. Math. Phys. {\bf 36}, 6377 (1995) 
[hep-th/9409089].
\bibitem{bb} J.D. Bekenstein, Phys. Rev. {\bf D23}, 287 (1981).
\bibitem{mal} J.M. Maldacena, Adv. Theor. Math. Phys. {\bf 2},
231 (1998) [hep-th/9711200].
\bibitem{gub} S.S. Gubser, I.R.  Klebanov and A.M. Polyakov,
Phys. Lett. {\bf B428}, 105 (1998)
 [hep-th/9802109].
\bibitem{wit} E. Witten, Adv. Theor. Math. Phys. {\bf 2}, 253 (1998)
[hep-th/9802150]; 
 {\it ibid}, 505 (1998) [hep-th/9803131].
\bibitem{agmoo} See, for a review, O. Aharony, S.S. Gubser, J. Maldacena,
H. Ooguri and  Y. Oz, Phys. Rept. {\bf 323}, 183 (2000) [hep-th/9905111].
\bibitem{sw} L. Susskind and E. Witten, ``The Holographic Bound in
Anti-de Sitter Space'', hep-th/9805114 (1998).
\bibitem{str2} A. Strominger, JHEP {\bf 0110}, 034 (2001) [hep-th/0106113].
\bibitem{med} S. Nojiri and S.D. Odintsov, ``Asymptotically
de Sitter Dilatonic Space-time, Holographic RG Flow and Conformal
Anomaly from (dilatonic) dS/CFT Correspondence'',  hep-th/0201210 (2002).
\bibitem{zmed} A.J.M. Medved, ``A Holographic Interpretation of
Asymptotically de Sitter Spacetimes'', hep-th/0112226 (2001).
\bibitem{zmyung} Y.S. Myung, ``Absorption Cross Section in de Sitter
Space'', hep-th/0201176 (2002).
\bibitem{zdas} S.R. Das, ``Thermality in de Sitter and Holography'',
hep-th/0202008 (2002).
\bibitem{deh} See, for instance,
 M. Spradlin, A. Strominger and A. Volovich, ``Les Houches
Lectures on de Sitter Space'', hep-th/0110007 (2001).
\bibitem{sussy} L. Susskind, ``Holography in the Flat Space Limit'',
hep-th/9901078 (1999).
\bibitem{ver} E. Verlinde, ``On the Holographic Principle in a
Radiation Dominated Universe'', hep-th/0008140 (2000).
\bibitem{sav} I. Savonije and E. Verlinde, Phys. Lett. {\bf B507},
305 (2001) [hep-th/0102042].
\bibitem{car} J.L. Cardy, Nucl. Phys. {\bf B270}, 317 (1986).
\bibitem{vir} P. Di Francesco, P. Mathieu and D. Senechal, {\it
Conformal Field Theory} (Springer, New York, 1997). 
\bibitem{cai} R.-G. Cai, Phys. Rev. {\bf D63}, 124018 (2001)
[hep-th/0102113].
\bibitem{dan} U.H. Danielsson, ``A Black Hole Hologram in de Sitter
Space'', hep-th/0110265 (2001).
\bibitem{klemm} D. Klemm, A.C. Petkou, G. Siopsis and D. Zanon,
Nucl. Phys. {\bf B620}, 519 (2002) [hep-th/0104141].
\bibitem{jing} J. Jing, ``Cardy-Verlinde Formula and Asymptotically
Flat Rotating Charged Black Holes'', hep-th/0202052 (2002).
\bibitem{youm} D. Youm, ``A Note on the Cardy-Verlinde Formula'',
hep-th/0201268 (2002).
\bibitem{zklemm} D. Klemm, A.C. Petkou and G. Siopsis,
Nucl. Phys. {\bf B601}, 380 (2001) [hep-th/0101076].
\bibitem{hal9000} E. Halyo, ``On the Cardy-Verlinde Formula and
the de Sitter/CFT Correspondence'', hep-th/0112093 (2001).
\bibitem{zamo} A.B. Zamolodchikov, JETP Lett. {\bf 43}, 730 (1986).
\bibitem{carlip} S.Carlip,  Phys. Rev. Lett. {\bf 82}, 2828 (1999)
[hep-th/9812013]; Class. Quant. Grav. {\bf 16}, 3327 (1999)
[gr-qc/9906126].
\bibitem{sol} S.N. Solodukhin, Phys. Lett. {\bf B454}, 213 (1999)
[hep-th/9812056].
\bibitem{crev} 
S. Carlip, Class. Quant. Grav. {\bf 15}, 3609 (1998)
[hep-th/9806026]; ``Black Hole Entropy from Horizon Conformal Field
Theory'', gr-qc/9912118 (1999).
\bibitem{xhal} E. Halyo, ``Strings and the Holographic Description
of Asymptotically de Sitter Spaces'', hep-th/0201174 (2002).
\bibitem{xsus} L. Susskind, ``Some Speculations about Black Hole Entropy
in String Theory'', hep-th/9309145 (1993).
\bibitem{xstu} L. Susskind, L. Thorlacius and J. Uglum, Phys. Rev. {\bf D48},
3743 (1993) [hep-th/9306069].
\bibitem{paper} A.J.M. Medved, ``AdS Holography
and Strings on the Horizon'', hep-th/0201215 (2002).
\bibitem{xtho} G. 't Hooft, Nucl. Phys. {\bf B335}, 138 (1990).
\bibitem{aaw} J. Attick and E. Witten, Nucl. Phys. {\bf B310}, 291 (1988).
\bibitem{sss} L. Susskind, ``Strings, Black Holes and Lorentz
Contraction'', hep-th/9308139 (1993).
\bibitem{xhrs} E. Halyo, A.Rajaraman and L. Susskind, Phys. Lett.
{\bf B392}, 319 (1997) [hep-th/9605112].
\bibitem{xhkrs} E. Halyo, B. Kol, A.Rajaraman and L. Susskind, Phys. Lett.
{\bf B401}, 15 (1997) [hep-th/9609075].
\bibitem{xhp} G.T.  Horowitz and J. Polchinski, Phys. Rev. {\bf D55},
 6189 (1997) [hep-th/9612146]; Phys. Rev. {\bf D57}, 2557 (1998)
[hep-th/9707170].
\bibitem{xhal2} E.Halyo, Int. J. Mod. Phys. {\bf A14}, 3831 (1999)
[hep-th/9610068]; Mod. Phys. Lett. {\bf A13}, 1521 (1998)
[hep-th/9611175];  ``De Sitter Entropy and Strings'',
hep-th/0107169 (2001); JHEP {\bf 0112}, 005 (2001) [hep-th/0108167].
\bibitem{xdv} T. Damour and G. Veneziano, Nucl. Phys. {\bf B568}, 93
(2000) [hep-th/9907030].
\bibitem{xdam} T. Damour, ``Quantum Strings and Black Holes'',
gr-qc/0104080 (2001).
\bibitem{xugk} U. Danielsson, A. Guijosa and M. Kruczenski, JHEP
{\bf 0109}, 011 (2001) [hep-th/0106201].
\bibitem{vaf} A. Strominger and C. Vafa, Phys. Lett. {\bf B379},
99 (1996) [hep-th/9601029].
\bibitem{xpeet} See, for instance,  
A.W. Peet, ``TASI Lectures on Black Holes in String
Theory'', hep-th/0008241 (2000).
\bibitem{xsen} A. Sen, Mod. Phys. Lett. {\bf A10}, 2081 (1995) 
[hep-th/9504147].
\bibitem{gh2} G.W. Gibbons and 
S.W. Hawking, Phys. Rev. {\bf D15}, 2752 (1977).
\bibitem{big} See, for instance, D. Bigatti and L. Susskind,
``TASI Lectures on the Holographic Principle'', hep-th/0002044 (2000).
\bibitem{bek} J.D. Bekenstein, Lett. Nuovo. Cim. {\bf 4}, 737 (1972);
Phys. Rev. {\bf D7}, 2333 (1973); Phys. Rev. {\bf D9}, 3292 (1974).
\bibitem{haw} S.W. Hawking, Comm. Math. Phys. {\bf 25}, 152 (1972);
 J.M. Bardeen, B. Carter and S.W. Hawking, Comm. Math. Phys.
{\bf 31}, 161 (1973).
\bibitem{bal} V. Balasubramanian and P. Kraus, Commun. Math. Phys. {\bf 208},
413 (1999) [hep-th/9902121].
\bibitem{newbal} V. Balasubramanian, J. de Boer and D. Minic,
``Mass, Entropy and Holography in Asymptotically de Sitter Spaces'',
hep-th/0110108 (2001).



\end{thebibliography}
\end{document}